\documentclass[nofootinbib,aps,prl,twocolumn,showpacs,superscriptaddress, reprint]{revtex4-2}

\usepackage{CJK}
\usepackage{graphicx}
\usepackage{dcolumn}
\usepackage{amsmath}
\usepackage{bm}
\usepackage{hyperref}
\usepackage{subfigure}
\usepackage{xcolor}
\usepackage{txfonts}
\usepackage{natbib}
\usepackage{booktabs}
\usepackage{mathtools}
\usepackage[mathlines]{lineno}

\renewcommand{\thefootnote}{\fnsymbol{footnote}}

\begin{document}

\preprint{APS/123-QED}


\title{Observation of a spectral hardening in cosmic ray boron spectrum\\ with the DAMPE space mission}

\noindent
\author{
 F.~Alemanno$^{1,2}$,
 C.~Altomare$^{3}$,
 Q.~An$^{4,5}$,
 P.~Azzarello$^{6}$, 
 F.~C.~T.~Barbato$^{7,8}$, 
 P.~Bernardini$^{1,2}$, 
 X.~J.~Bi$^{9,10}$,
 H.~Boutin$^{6}$,
 I.~Cagnoli$^{7,8}$,
 M.~S.~Cai$^{11,12}$
% ...
 \\
 (DAMPE Collaboration)*\footnotemark[1]
 \\
\noindent
$^1$Dipartimento di Matematica e Fisica E. De Giorgi, Università del Salento, I-73100, Lecce, Italy \\
$^2$Istituto Nazionale di Fisica Nucleare (INFN) - Sezione di Lecce, I-73100, Lecce, Italy \\
$^3$Istituto Nazionale di Fisica Nucleare (INFN) - Sezione di Bari, I-70125, Bari, Italy \\
$^4$State Key Laboratory of Particle Detection and Electronics, University of Science and Technology of China, Hefei 230026, China \\
$^5$Department of Modern Physics, University of Science and Technology of China, Hefei 230026, China \\
$^6$Department of Nuclear and Particle Physics, University of Geneva, CH-1211 Geneva, Switzerland \\ 
$^7$Gran Sasso Science Institute (GSSI), Via Iacobucci 2, I-67100 L’Aquila, Italy \\
$^8$Istituto Nazionale di Fisica Nucleare (INFN) - Laboratori Nazionali del Gran Sasso, I-67100 Assergi, L’Aquila, Italy \\
$^9$Institute of High Energy Physics, Chinese Academy of Sciences, Yuquan Road 19B, Beijing 100049, China \\
$^{10}$University of Chinese Academy of Sciences, Yuquan Road 19A, Beijing 100049, China \\
$^{11}$Key Laboratory of Dark Matter and Space Astronomy, Purple Mountain Observatory, Chinese Academy of Sciences, Nanjing 210023, China \\
$^{12}$School of Astronomy and Space Science, University of Science and Technology of China, Hefei 230026, China \\
$^{13}$Istituto Nazionale di Fisica Nucleare (INFN) - Sezione di Perugia, I-06123 Perugia, Italy \\
$^{14}$Institute of Modern Physics, Chinese Academy of Sciences, Nanchang Road 509, Lanzhou 730000, China \\ 
$^{15}$National Space Science Center, Chinese Academy of Sciences, Nanertiao 1, Zhongguancun, Haidian district, Beijing 100190, China \\
$^{16}$Dipartimento di Fisica “M. Merlin” dell’Università e del Politecnico di Bari, I-70126, Bari, Italy \\
$^{17}$Department of Physics and Laboratory for Space Research, the University of Hong Kong, Pok Fu Lam, Hong Kong SAR, China
}
\date{\today}

\begin{abstract}
Secondary cosmic ray fluxes are important probes of the propagation and interaction of high-energy particles in the Galaxy. Recent measurements of primary and secondary cosmic ray nuclei have revealed unexpected spectral features that demand a deeper understanding. In this work we report the direct measurement of the cosmic ray boron spectrum from 10 GeV/n to 8 TeV/n with eight years of data collected by the Dark Matter Particle Explorer (DAMPE) mission. The measured spectrum shows an evident hardening at $182\pm24$ GeV/n with a spectral power index of $\gamma_1 = 3.02 \pm 0.01$ before the break and an index change of $\Delta \gamma = 0.31 \pm 0.05$ after the break. A simple power law model is disfavored at a confidence level of 8$\sigma$. Compared with the hardenings measured in the DAMPE proton and helium spectra, the secondary boron spectrum hardens roughly twice as much as these primaries, which is consistent with a propagation related mechanism to interpret the spectral hardenings of cosmic rays observed at hundreds of GeV/n.
\end{abstract}

\maketitle
\normalsize

\renewcommand{\thefootnote}{\arabic{footnote}}
\newcommand{\note}[1]{{\bf\color{red} #1}}

\emph{\label{sec:intro}Introduction} - The origin and transport of Galactic Cosmic Rays 
(GCRs) --- high energy particles that are accelerated and travel through our Galaxy ---
and their interplay with the interstellar medium (ISM) are of great importance in
astrophysics \cite{Aloisio2018}. The lithium, beryllium, and boron in GCRs, with 
abundances that are orders of magnitude higher than that expected from stellar nucleosynthesis, 
are mainly produced by spallation of heavier nuclei with the ISM. These nuclei are therefore
called secondary cosmic rays, which offer a unique probe of the propagation and 
interaction processes of GCRs. The spectra of GCRs show hardenings at a few hundred 
GeV per nucleon, as reported by many experiments for virtually all primary species
such as protons, He, C, O, Ne, Mg, Si, Fe
\cite{p_he_cream, pamela_proton, Panov2009EnergySO, ADRIANI2013219, proton_ams, helium_ams, 
c_o_ams, nemgsi_ams, fe_ams, p_dampe, he_dampe, p_calet, he_calet}, and becomes softer 
again at around 10 TeV per nucleon as recently observed in the proton and helium spectra 
\cite{p_dampe, he_dampe, p_calet, he_calet,Atkin:2018wsp}. Recent measurements of the boron 
spectrum by AMS-02 in the 1.9 GV $-$ 3.3 TV rigidity range \cite{libeb_ams} and by CALET 
in the energy range of 8.4 GeV/n to 3.8 TeV/n \cite{b_calet} also gave hints of a hardening at $E_{\rm br} \sim 200 $ GeV/n. However, with the available statistics, the significance of the hardening is not high. The spectral behavior goes beyond a single power law, demanding a deeper understanding of the acceleration and/or propagation mechanisms of GCRs and highlighting the necessity of precise measurements extending to even higher energies. 

The Dark Matter Particle Explorer (DAMPE) is a space-based GCR and $\gamma$-ray detector 
that has been smoothly operating since its launch in December, 2015 \cite{dampe_detector}. 
The instrument includes four main sub-detectors: a plastic scintillator (PSD) designed to 
measure the absolute value of the charge of the incoming particles and to act as a veto for gamma-ray identification \cite{dampe_psd}; a silicon-tungsten tracker-converter (STK) to measure particle direction and charge, and convert impinging gamma rays into electron-positron pairs \cite{dampe_stk}; 
a thick bismuth germanium oxide electromagnetic calorimeter (BGO) to measure the energy 
and direction of particles and to distinguish hadronic and electromagnetic showers 
\cite{dampe_bgo}; a neutron detector (NUD) to further enhance the discrimination between 
hadronic and electromagnetic particles \cite{dampe_nud}. 
DAMPE is calibrated with the on-orbit data
\cite{dampe_calib,dampe_psd_cal,dampe_stk_cal,dampe_bgo_cal,dampe_nud_cal}, enabling 
precise measurements of various kinds of particles in a wide energy range.
Here we present the measurement of the cosmic ray boron fluxes in an energy range from 
$\hbox{10 GeV/n}$ to $\hbox{8 TeV/n}$, using 8 years of DAMPE data. 

\emph{\label{sec:MC}Monte Carlo simulations} -- Monte Carlo (MC) simulations using the
GEANT4 toolkit (version 4.10.5) \cite{geant4} are performed to study the response of the 
detector to different particles in a wide energy range. For boron nuclei with primary energies 
between 10 GeV and 100 TeV, we employ the FTFP$\_$BERT physics list, and for energy above 
100 TeV, the EPOS-LHC model \cite{epos_lhc} which is linked to the GEANT4 with the Cosmic 
Ray Monte Carlo (CRMC) package \footnote{\url{https://web.ikp.kit.edu/rulrich/crmc.html}} 
\cite{andrii_crmc}. The events are initially generated uniformly 
from a hemispherical surface around the detector with isotropic direction, following a $E^{-1}$
spectrum. Simulated events are then re-weighted during the analysis to a $E^{-3}$ spectrum. 
Both $^{10}$B an $^{11}$B isotopes are simulated, and the isotopic ratio can be re-weighted 
in the analysis. 
The systematic uncertainties related to the spectral shape and the isotopic ratio applied to re-weighting the MC simulations are evaluated carefully and included in the final spectral measurement. 
Additional MC simulations are produced 
with FLUKA 2011.2x \cite{fluka} in order to evaluate the uncertainty related to the hadronic 
interaction models. For the FLUKA simulation of boron, the deposited energy response in the BGO calorimeter is corrected by a global factor of 0.95 for all energies based on the comparison with beam-test data \cite{Chen2023}. While, no correction is performed for the GEANT4 simulaiton as the energy response is consistent with beam-test data within $\sim$2\% \cite{Chen2023}.

\emph{\label{sec:selection} Event Selection} - In this analysis, 8 years of on-orbit data are used, recorded by DAMPE between January 1, 2016 and December 31, 203. The events collected 
when the satellite crosses the South Atlantic Anomaly (SAA) region are excluded. After 
subtracting the instrument's dead time ($\sim$18.44\% of the operation time), the time 
dedicated to on-orbit calibration ($\sim$1.56\%), a giant solar flare between September 9 
and September 13, 2017 \footnote{\url{https://solarflare.njit.edu/datasources.html}}, 
and the SAA passage time ($\sim$5\%) \cite{dampe_calib}, the total exposure time is 
$\sim$1.4$\times 10^{8}$ s, corresponding to $\sim$73.9\% of the operation time. 
A general pre-selection is then applied to select events with good quality. A selection
tailored to the boron nuclei follows.   

\emph{(i) Pre-selection} is a set of requirements that ensure a 
proper event reconstruction and shower development, mainly involving the BGO calorimeter. 
They include: 
\begin{itemize}
    \item the total deposited energy in the BGO calorimeter must be larger than 40 GeV, corresponding to an incident energy larger than  $\sim$80 GeV for boron nuclei, to avoid the geomagnetic rigidity cutoff effect \cite{geomag_eff};
    \item the energy deposit in each BGO layer must be lower than 35\% of the total deposited energy in the calorimeter, to reject events entering from the side; 
    \item the BGO bar with the highest 
    energy deposit in the first four layers must not be on the edge, to ensure a good
    containment of the shower in the calorimeter.  
\end{itemize}

\emph{(ii) Trigger selection} - Events passing the High Energy Trigger (HET) are selected. 
The requested condition is a deposited energy larger than $\sim10$ MIPs in the first three 
BGO layers and larger than $\sim2$ MIPs in the fourth layer (one MIP releases $\simeq$ 23 
MeV in a BGO bar) \cite{Zhang_2019}. The HET is designed to select good electromagnetic shower events. For boron nuclei, the HET condition is easily satisfied and the efficiency is very high (see the {\tt Supplemental Material}).

\emph{(iii) Track selection} - The particle trajectory of each event is obtained by by selecting the one track from multiple candidates STK tracks reconstructed with the Kalman filter algorithm \cite{stk_track}. The good candidate tracks are confined as that the number of hits on the track is larger than 3 in both $xz$ and $yz$ layers, the $\chi^2$/d.o.f. value of the Kalman filter is smaller than 15, the
angular deviation from the BGO shower axis is within $25^{\circ}$ and the average distance with the energy CoG (center of gravity) positions in the first four BGO layers is within 15 mm. 
Then, the track with the maximum average hit energy is chosen as the best one. The selected track is further required to match the maximum PSD hits in both PSD-x and PSD-y layers, and pass through the calorimeter from top to bottom. 
Apart from the Kalman filter track reconstruction, the machine learning track approach \cite{stk_MLtrk} is also applied as a cross-check in this analysis. These two different track reconstruction methods show a very good consistency on the boron spectral measurement. 

\emph{(iv) Charge selection} - The particle charge $Z$ is reconstructed with the ionization 
energy deposited in both PSD and STK. The measured charge value from the hit of the first 
STK layer along the track is set to be larger than 3 ($Q_{\rm STK1}>3$), in order to suppress
protons and helium events. The PSD is composed of four sub-layers placed in a hodoscopic 
configuration in $yz$-view and $xz$-view, which provides, at most, four independent charge 
measurements. A global PSD charge is defined by considering the PSD signals in a different
set of the 4 sub-layers to reduce the mis-identification for events undergoing inelastic
interactions inside the PSD (see the {\tt Supplemental Material}).
Boron candidates are selected in the charge interval that varies with the deposited energy($E_{\rm dep}$) as follows: $4.7 < Q_{\rm PSD} < 5.25 + 0.01\cdot \log^2(E_{\rm dep}/{\rm GeV})$. Such an energy dependence is introduced in order to maintain a uniform charge selection efficiency of $\sim$90\% in the entire energy range of the analysis. In total, $8.25 \times 10^{5}$ boron candidates with $E_{\rm dep} > 40$ GeV are selected for the spectral measurement.

\begin{figure}[htpb]
\centering
\includegraphics[width=0.48\textwidth]{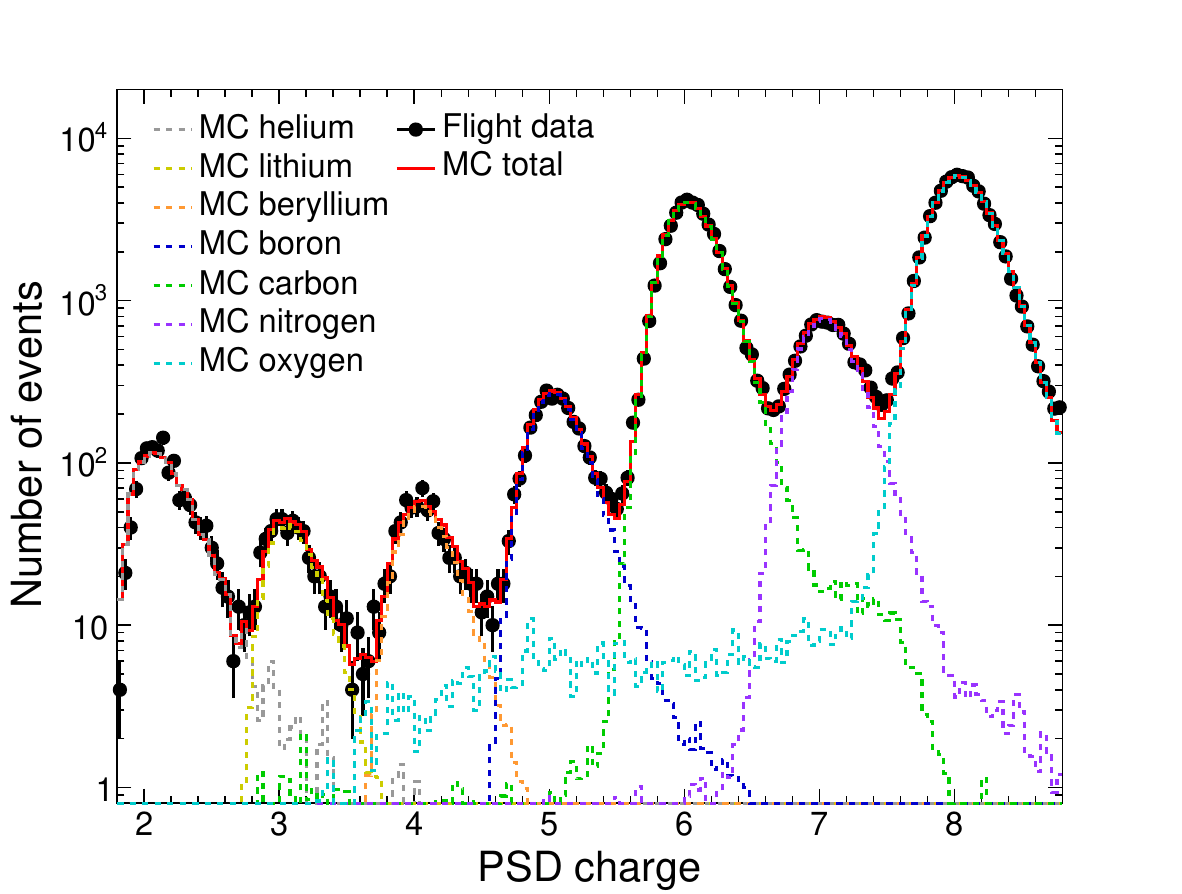}
\includegraphics[width=0.48\textwidth]{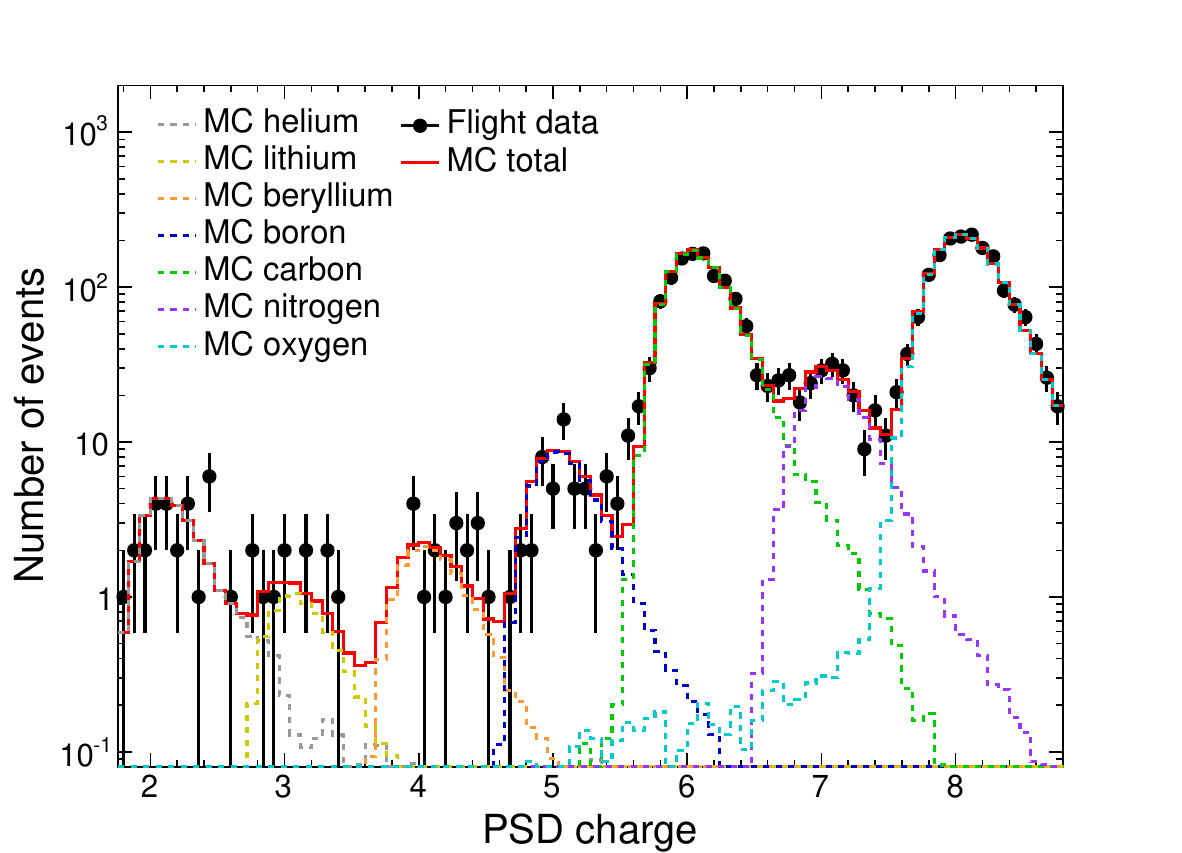}
\caption{Distributions of $Q_{\rm PSD}$ for flight data (black points) in two energy bins 
of $E_{\rm dep}$: [0.63, 2] TeV (top), and [6.3, 20] TeV (bottom). The best fit of the 
total MC templates is shown as the red histogram, while the templates of each nucleus are 
shown in different color according to the legend.}
\label{fig:MC_templatefit}
\end{figure}

\emph{(v) Background subtraction} - Due to its secondary origin, boron has a 
relatively low flux with respect to abundant primaries. For this reason the background
contamination from different nuclear species has to be carefully evaluated. A MC-based 
template fit is performed on the $Q_{\rm PSD}$ variable to estimate the background in 
the selected PSD charge window of boron, as shown in Fig.~\ref{fig:MC_templatefit}. 
The MC simulations for nuclei from helium to oxygen, generated with GEANT4, are shown 
by dashed lines. 
The red solid line corresponding to the sum of all the species shows a good fit to the flight data.
The background from fluorine and heavier nuclei is neglected in this analysis, as their fluxes are much lower than those of carbon and oxygen. 
The estimated contamination fractions from different species are shown in
Fig.~S1 of the {\tt Supplemental Material}. The most prominent sources of background 
are carbon and oxygen, especially at lower energies. The contributions from helium and 
beryllium become significant with increasing energy. The overall estimated background is 
found to be 1\% to 2\% for $E_{\rm dep}<100$ GeV and $\sim$5\% around 10 TeV. Even at the 
highest deposited energies, the background fraction is small enough, $\sim$6\%, ensuring a pure sample for the flux measurement.

\emph{Effective acceptance} - The effective acceptance in the $i$-th bin of incident kinetic 
energy can be estimated from the MC samples as $A_i(E_{\rm inc}^i) = G_{\rm gen} \cdot N_{\text{sel}}^i/N_{\text{gen}}^i$, where $G_{\rm gen}$ is the geometrical factor 
corresponding to the surface used to generate events in the MC simulation, $N_{\text{gen}}^i$ 
is the number of generated events in the $i$-th energy bin, and $N_{\text{sel}}^i$ is the 
number of events that pass the selection. The effective acceptance as a function of incident kinetic energy is shown in Fig. S2 of 
the {\tt Supplemental Material}.     

\emph{Energy measurement \& unfolding procedure} - The energy is measured through adding
the energy recorded by the BGO bars together. If the energy deposited in a single BGO bar 
exceeds $\sim4$ TeV, some readout channels may saturate, leading to non-linearity in the 
energy response. A saturation correction based on MC simulation \cite{dampe_saturation} 
was developed to correct the energy measurement when one or more bars are saturated. 
Another effect to be taken into account is the signal quenching in the BGO bars \cite{Wei2020}.
This effect is corrected based on comparing MC samples with the quenching switched on and off in the simulation of ionization energy deposits in the calorimeter \cite{Chen2023}. 
The quenching effect for boron nuclei results in about 2\% lower energy deposit in the 
BGO for a kinetic energy of 200 GeV.

Since the thickness of the BGO calorimeter is of $\sim$ 1.6 nuclear interaction lengths, it is not able to fully contain the hadronic showers developed by CR nuclei. The energy deposited in the calorimeter will only be a fraction of its primary energy ($\sim 35\%-45\%$), subject to large fluctuations. For this reason it is difficult to extract the incident kinetic energy $E_{\rm inc}$ from the deposited energy $E_{\rm dep}$ event by event, and a statistical unfolding procedure is needed. The number of events in the $i$-th bin of BGO deposited energy, $N(E_{\rm dep}^i)$, is related to the number of events in the $j$-th bin of incident kinetic energy $N(E_{\rm inc}^j)$ by
\begin{equation}
N(E_{\rm dep}^i)(1-\beta_{i}) = \sum_{j=1}^{n}P\small(E^i_{\rm dep} | E^j_{\rm inc}\small) N(E_{\rm inc}^j),
\label{Formula:unfolding}
\end{equation}
where $\beta_{i}$ is the background fraction, $P\small(E^i_{\rm inc} | E^j_{\rm dep}\small)$ is the energy response matrix which is derived using 
MC simulations. An iterative Bayesian unfolding approach with the smoothing regularization \cite{DAGOSTINI2010} is adopted to calculate the number of events in each $E_{\rm inc}$
bin. 

\emph{Flux calculation} - The differential flux for each energy bin $\Phi_i$ as a function 
of the incident kinetic energy can be computed as 
\begin{equation}
    \Phi_i = \Phi(E_{\rm inc}^i, E_{\rm inc}^i + \Delta E_i)=\frac{N(E_{\rm inc}^i)}
    {\Delta T \cdot A_i \cdot \Delta E_i},
\end{equation}
where $\Delta E_i$ is the width of the $i$-th kinetic energy bin, $N(E_{\rm inc}^i)$ is the 
number of events in the $i$-th kinetic energy bin after the unfolding, $\Delta T$ is the 
total exposure time, and $A_i$ is the effective acceptance. The fluxes are then converted 
to kinetic energy per nucleon assuming the isotopic ratio of 
$Y=^{11}{\rm B}/(^{10}{\rm B}+^{11}{\rm B}) = 0.7$ \cite{libeb_ams}.

\emph{Uncertainty analysis} - The statistical uncertainties are associated with Poissonian 
fluctuations of the number of detected events $N(E_{\rm dep}^j)$. However, since the unfolding
process introduces a bin-to-bin migration, this uncertainty cannot be directly translated to 
the kinetic energy bins. To properly determine the error propagation in the unfolding procedure,
a toy-MC approach is applied, sampling the event numbers in each deposited energy bin with Poisson fluctuations. The variations of the unfolded numbers of events in each kinetic energy bin are then obtained. The standard deviation of the resulting boron flux distribution in each bin is assigned as the 1$\sigma$ statistical error.

The systematic uncertainties due to different sources are investigated extensively, 
including the selection efficiencies, the unfolding procedure, the background subtraction, 
the isotope composition of boron, and the hadronic model. The selection efficiencies of the simulation are compared with the flight data and the deviations are taken as the related systematic uncertainties. The agreement between simulation and data is within 1.5\% for the HET efficiency, 3\% for the STK track efficiency, 2\% for the STK charge efficiency (see the {\tt Supplemental Material}). The uncertainty 
of the PSD charge selection is estimated by varying the PSD charge selection window 
to change the efficiency by $-15\%$ to $+5\%$ and checking the flux variations. 
The uncertainty due to the background subtraction is estimated by considering the errors of the contamination fractions (see the {\tt Supplemental Material}). The uncertainty of the unfolding procedure, related to the initial value of the spectral index and the finite statistics of the 
MC sample, is estimated to affect the flux measurement by (0.5-1.0)\%. 
The uncertainty due to the assumed boron isotope composition of $Y=0.7\pm0.1$ 
\cite{libeb_ams} is estimated to be $\sim1.9$\%. Finally, the systematic uncertainty 
associated with the hadronic model used in the MC simulation is obtained through 
comparing simulation results with the test beam data or two different simulation tools. 
Specifically, below 75 GeV/n an uncertainty of 4\% is assigned according to the 
comparison of the energy response between the test beam data and the GEANT4 simulation. At higher energies, the uncertainty is assigned as the difference between the fluxes measured using the GEANT4 and the FLUKA simulations. The difference becomes larger with 
the increase of energy and reaches $\sim20$\% at the highest energies. The statistical 
and systematic uncertainties for different incident energies are summarized in Fig. S7 
of the {\tt Supplemental Material}.

\begin{figure}[htbp]
\centering
\includegraphics[width=0.45\textwidth]{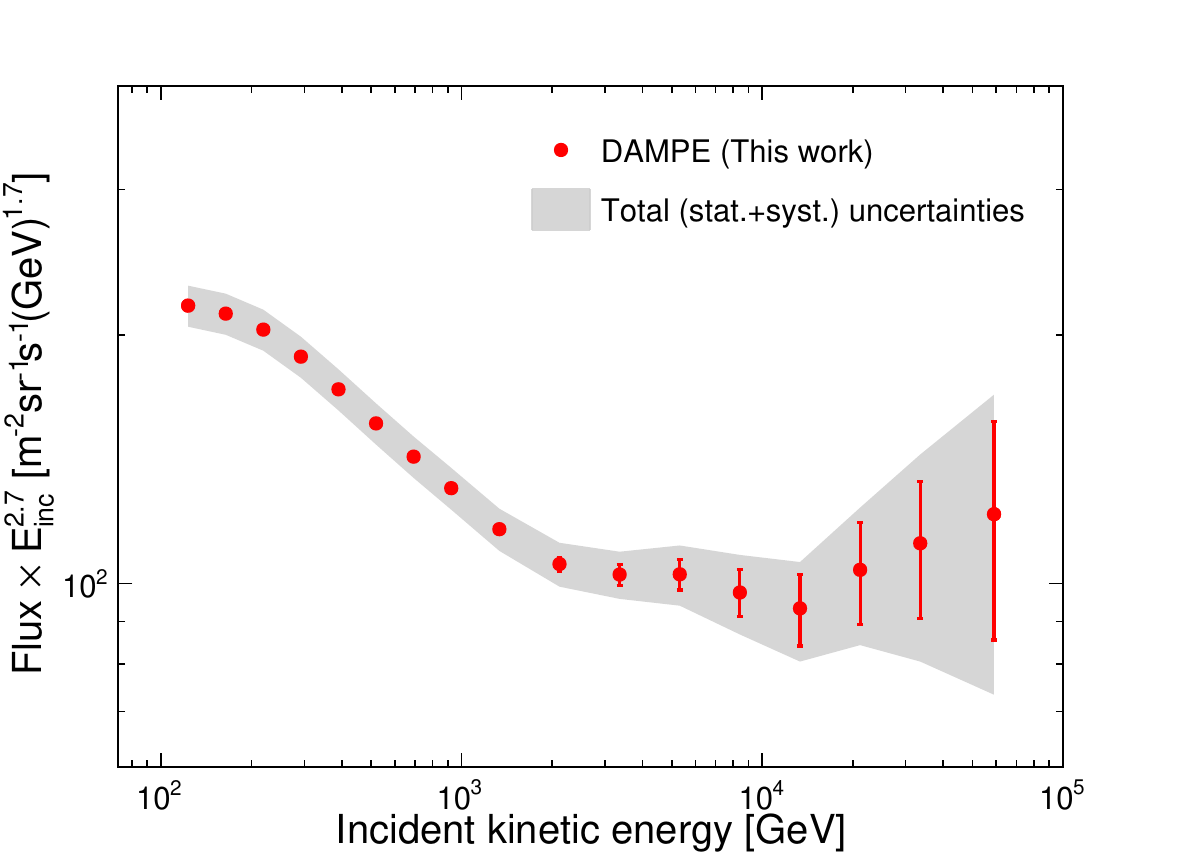}
\includegraphics[width=0.45\textwidth]{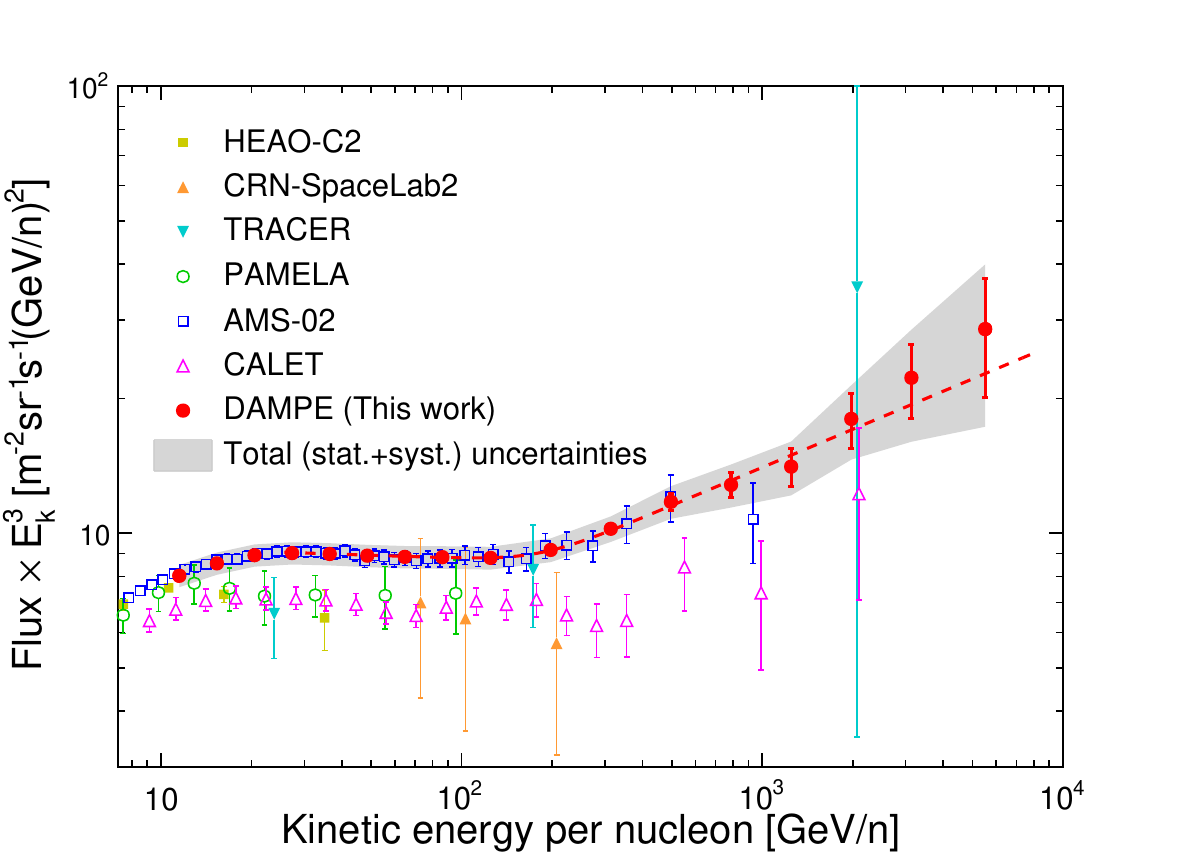}
\caption{Top panel: boron spectrum weighted by $E^{2.7}$ as measured with DAMPE. 
The statistical errors are represented by red error bars and the total errors, given by 
the sum in quadrature of statistical and systematic errors, are shown by the gray shaded area. 
Bottom panel: comparison of the DAMPE spectrum (converted to kinetic energy per nucleon 
assuming the isotopic composition of $Y=0.7$ \cite{libeb_ams} and weighted by $E^{3}$) with previous measurements by HEAO-C2 \cite{Engelmann_1990}, CRN-SpaceLab2 \cite{Swordy_1990}, TRACER \cite{Obermeier_2011}, PAMELA \cite{Adriani_2014}, AMS-02 \cite{PhysRevLett.130.211002}, and CALET \cite{calet_boron}. For the AMS-02 results \cite{PhysRevLett.130.211002}, the fluxes are converted from rigidity to kinetic energy per nucleon assuming the isotopic composition of $Y=0.7$.}
\label{fig:b_flux}
\end{figure}

\emph{Results} - The boron spectrum measured with DAMPE is shown in the top panel of 
Fig.~\ref{fig:b_flux}. The spectrum converted to kinetic energy per nucleon, assuming 
the isotopic composition of $Y=0.7$, is shown in the bottom panel. The corresponding 
data tables including statistical and systematic errors are reported in the 
{\tt Supplemental Material}. The DAMPE spectrum is compared with previous measurements 
by PAMELA \cite{Adriani_2014}, AMS-02 \cite{PhysRevLett.130.211002}, and CALET 
\cite{calet_boron}. At low energies, our result is in agreement, within uncertainties, with the AMS-02 
and PAMELA measurements. 
Compared with other measurements, the DAMPE result extends to higher energies and exhibits a clear hardening above $O(100)$ GeV/n. To quantitatively obtain the properties of the hardening, 
the spectrum is fitted by a Smoothly Broken Power Law (SBPL) function
\begin{equation}
    \Phi_{\rm SBPL}(E_k) = \Phi_0 \cdot \left( \frac{E_k}{\text{GeV/n}} \right)^{-\gamma_1} 
    \left[1 + \left(\frac{E_k}{E_{\rm br}}\right)^s \right]^{\frac{\Delta \gamma}{s}}.
\label{eq:sbpl}
\end{equation}
The dashed line shown in Fig.~\ref{fig:b_flux} represents the best SBPL fit in the energy 
range from 25 GeV/n to 8 TeV/n. The fit gives $\chi^2$/d.o.f.$=1.7/10$. The hardening is 
identified at $E_{\rm br} = 182 \pm 24$ GeV/n with a slope before the break of 
$\gamma_1 = 3.02 \pm 0.01$ and an index change of $\Delta \gamma = 0.31 \pm 0.05$. 
The significance of the hardening, when tested against a single power law (PL) fit 
which gives $\chi^2$/d.o.f.$=71.3/12$, is $8\sigma$. More details about the spectral 
fit can be found in the {\tt Supplemental Material}.  

\begin{figure}[]
    \centering
    \includegraphics[width=0.45\textwidth]{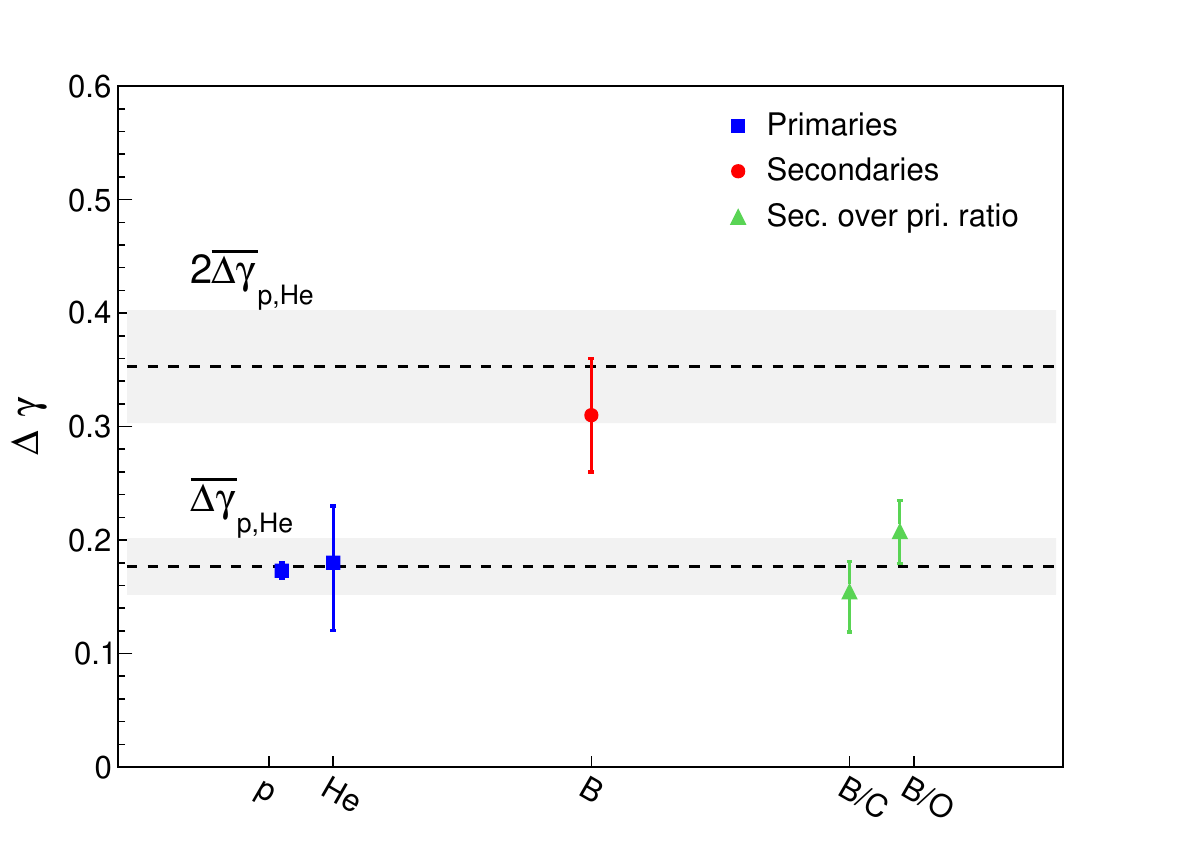}
    \caption{The $\Delta\gamma$ values of the hardening in the spectra of proton, helium, B/C, B/O and boron measured with DAMPE.}
    \label{fig:dgamma}
\end{figure}

\emph{Conclusion and discussion} - 
The measurement of the cosmic ray boron spectrum with 8 years of DAMPE data is presented 
in this work. The flux was measured in a kinetic energy per nucleon range from 10 GeV/n 
to 8 TeV/n. Compared with previous measurements, the DAMPE spectrum presents a much 
improved precision at high energies ($>1$ TeV/n). 
For the first time, the DAMPE measurement provides an observation of spectral hardening at 
$E_{\rm br} = 182 \pm 24$ GeV/n, with a significance of 8$\sigma$. The spectral index 
below the hardening energy is $3.02 \pm 0.01$, and the change of the spectral index above 
$E_{\rm br}$ is $0.31 \pm 0.05$. 
Previous measurements by AMS-02 \cite{PhysRevLett.130.211002} and CALET \cite{calet_boron} suggested such hardening with limited statistical significance, without claiming a definitive conclusion, due to the smaller statistics and lower energy reach. The spectral change of $\Delta \gamma = 0.31 \pm 0.05$ suggests that secondary boron hardens about twice as much as primaries such as proton and helium as previously 
measured by DAMPE \cite{p_dampe,he_dampe}, as shown in Fig.~\ref{fig:dgamma}. The result is also consistent with the hardenings 
in boron-to-carbon and boron-to-oxygen flux ratios \cite{bcbo_dampe}, if carbon and oxygen 
show similar hardenings with protons and helium. Such an observational fact supports the propagation mechanism to account for spectral
hardenings of GCRs. This is because for a diffusion coefficient $\propto E^{-\delta}$ and 
an initial primary cosmic ray spectrum of $E^{-\alpha}$, the propagated primary 
(secondary) cosmic ray spectrum should be $E^{-\alpha-\delta}$ ($E^{-\alpha-2\delta}$). 
As a consequence, if $\delta$ has a change of $\Delta\delta$, secondaries would show a
spectral change of $2\Delta\delta$, with respect to $\Delta\delta$ for primaries. Other mechanisms such as secondary production around sources or re-acceleration by nearby shocks may also explain the data \cite{Ma:2022iji,Malkov:2021gxd}.

\emph{Acknowledgments} - The DAMPE mission was funded by the strategic priority science and technology projects in space science of Chinese Academy of Sciences. In China the data analysis is supported by the National Key Research and Development Program of China (No. 2022YFF0503302), the National Natural Science Foundation of China (Nos. 12220101003, 11921003, 11903084, 12003076, 12022503 and 12103094), the Project for Young Scientists in Basic Research of the Chinese Academy of Sciences (No. YSBR-061), the Strategic Priority Program on Space Science of Chinese Academy of Sciences (No. E02212A02S),the Youth Innovation Promotion Association of CAS, the Young Elite Scientists Sponsorship Program by CAST (No. YESS20220197), the New Cornerstone Science Foundation through the XPLORER PRIZE and the Program for Innovative Talents and Entrepreneur in Jiangsu. In Europe the activities and data analysis are supported by the Swiss National Science Foundation (SNSF), Switzerland, the National Institute for Nuclear Physics (INFN), Italy, and the European Research Council (ERC) under the European Union’s Horizon 2020 research and innovation programme (No. 851103).

\renewcommand{\thefootnote}{\fnsymbol{footnote}}

\footnotetext[1]{dampe@pmo.ac.cn}

\footnotetext[2]{Now at Institute of Physics, Ecole Polytechnique F\'{e}d\'{e}rale de Lausanne (EPFL), CH-1015 Lausanne, Switzerland.}


\footnotetext[3]{Now at Dipartimento di Fisica e Chimica “E. Segrè”, Università degli Studi di Palermo, via delle Scienze ed. 17, I-90128 Palermo, Italy.}

\footnotetext[4]{Now at Inter-university Institute for High Energies, Université Libre de Bruxelles, B-1050 Brussels, Belgium}

\footnotetext[5]{Now at New York University Abu Dhabi, Saadiyat Island, Abu Dhabi 129188, United Arab Emirates.}

\footnotetext[6]{Also at School of Physics and Electronic Engineering, Linyi University, Linyi 276000, China.}

\bibliographystyle{apsrev4-2}
\bibliography{apssamp}

\clearpage

\setcounter{figure}{0}
\renewcommand\thefigure{S\arabic{figure}}
\setcounter{table}{0}
\renewcommand\thetable{S\arabic{table}}

\section{Supplementary Material}

\subsection{The global charge of PSD}
\begin{figure}[htpb]
    \centering
    \includegraphics[width=0.45\textwidth]{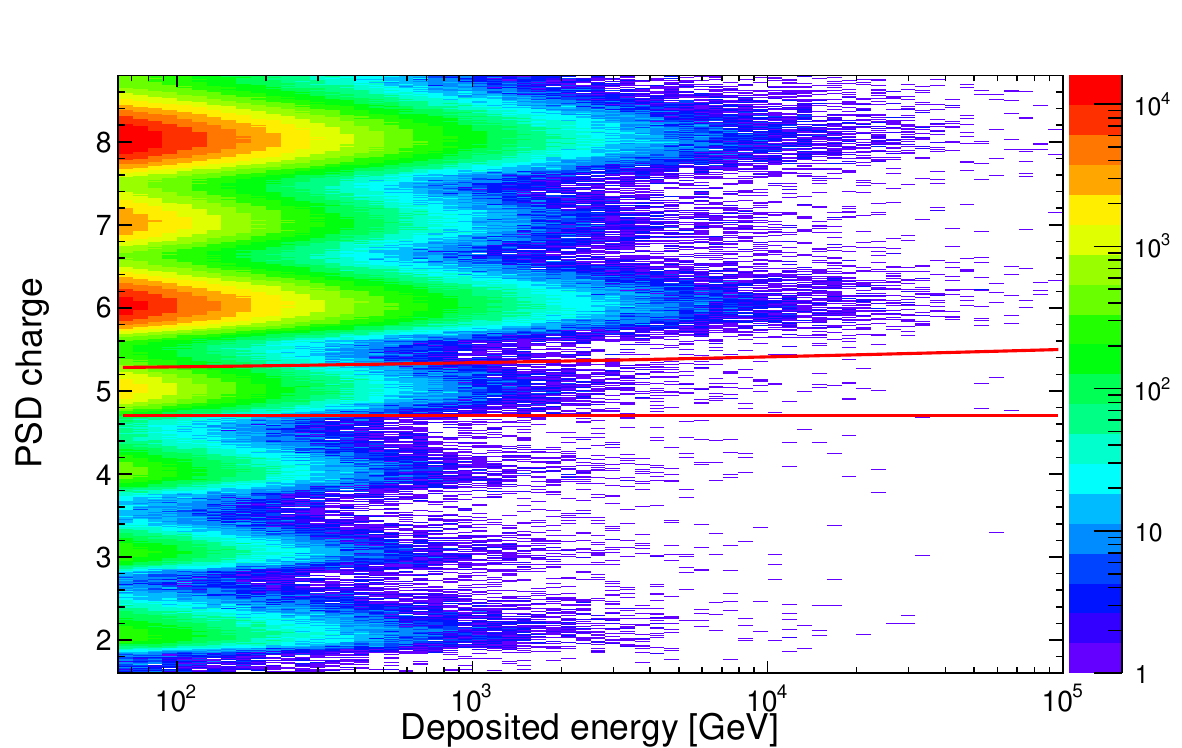}
    \caption{The PSD charge as a function of deposited energy for particles with $Z = 2 - 8$ 
    in the flight data. The red lines indicate the selection conditions of boron candidates.}
    \label{fig:psdQ-bgoE}
\end{figure}

The PSD is made of four sub-layers, two oriented in the $y$ direction and two in the $x$ direction, which provides at most four independent charge measurements. The signal of each PSD hit is corrected taking into account the light attenuation, the incident angle, quenching in the scintillator bars and equalization of different PMTs \cite{DONG201931,Ma_2019}. The global PSD charge estimator is defined as
\begin{equation}
Q_{\rm PSD} = \frac{\sum_{i} Q_{{\rm PSD},i}}{N_{\rm PSD}},
\label{Formula:PSDcharge}
\end{equation}
where the index $i$ goes over the PSD sub-layers with non-zero signal, checking that charges 
$Q_{{\rm PSD},i}$ in successive sub-layers satisfy the condition: 
\begin{equation}
|Q_{{\rm PSD},i} - Q_{{\rm PSD},i+1}| < 1,
\label{Formula:PSDdiff}
\end{equation}
and $N_{\rm PSD}$ is the number of sub-layers passing this requirement. The same procedure is applied to the MC simulations, and the MC charge distributions are shrinked to match the flight data.
The PSD charge as a function of deposited energy in the flight data is shown in Fig.~\ref{fig:psdQ-bgoE}. The boron candidates are selected with a deposited-energy-dependent charge interval (as shown by solid lines in Fig.~\ref{fig:psdQ-bgoE}) to keep a stable charge efficiency of $\sim$90\% in the whole energy range. 

\subsection{Background contamination}
\begin{figure}[htpb]
    \centering
    \includegraphics[width=0.45\textwidth]{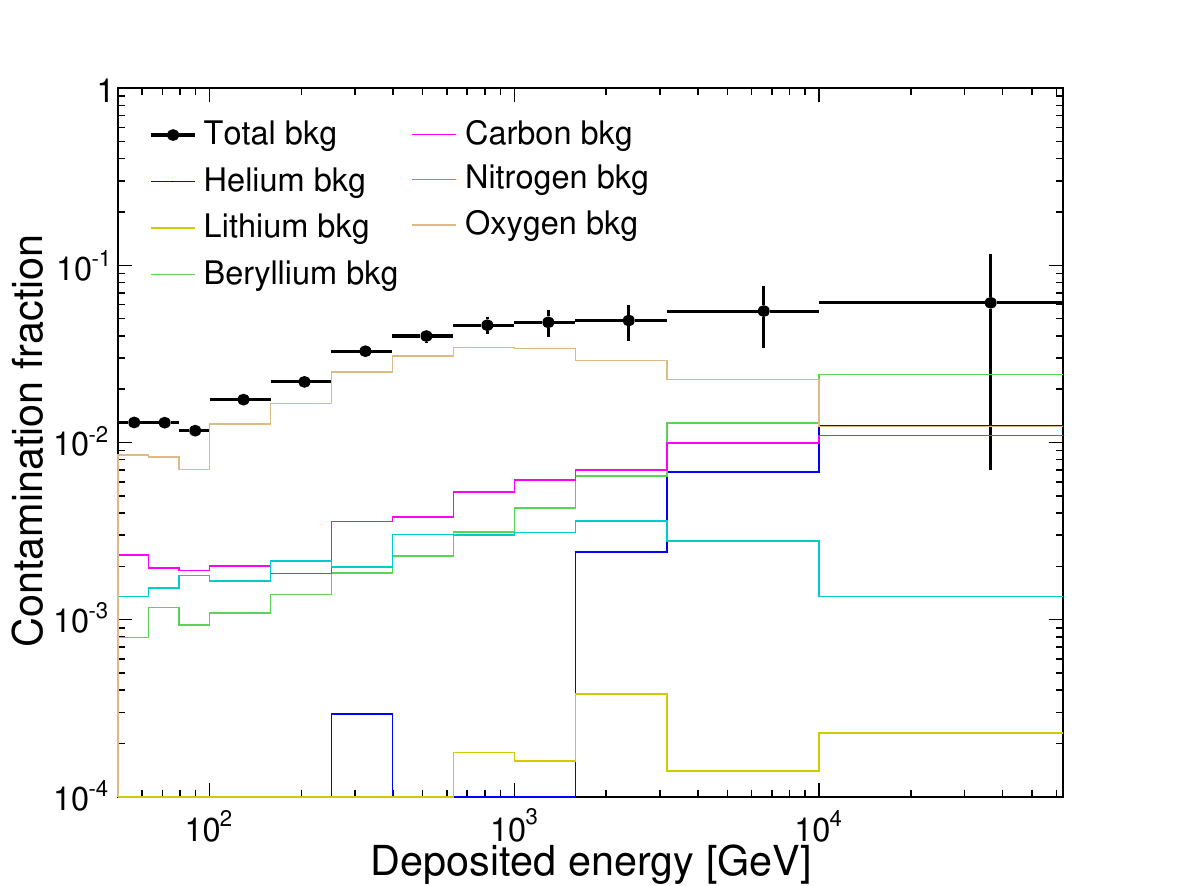}
    \caption{Background contamination fractions estimated with the MC template fit for boron.}
    \label{fig:bkg}
\end{figure}

Fig.~\ref{fig:bkg} shows the estimated fractions of background contamination for boron 
candidates in the selected PSD charge window obtained via the MC template fit, from helium, 
lithium, beryllium, carbon, nitrogen, and oxygen nuclei. The uncertainties of the total contamination fractions are estimated on the basis of Binomial distribution by considering the statistics of the selected boron candidates in each deposited energy bin. 

\subsection{Effective acceptance validation}
\begin{figure}[htpb]
    \centering
    \includegraphics[width=0.45\textwidth]{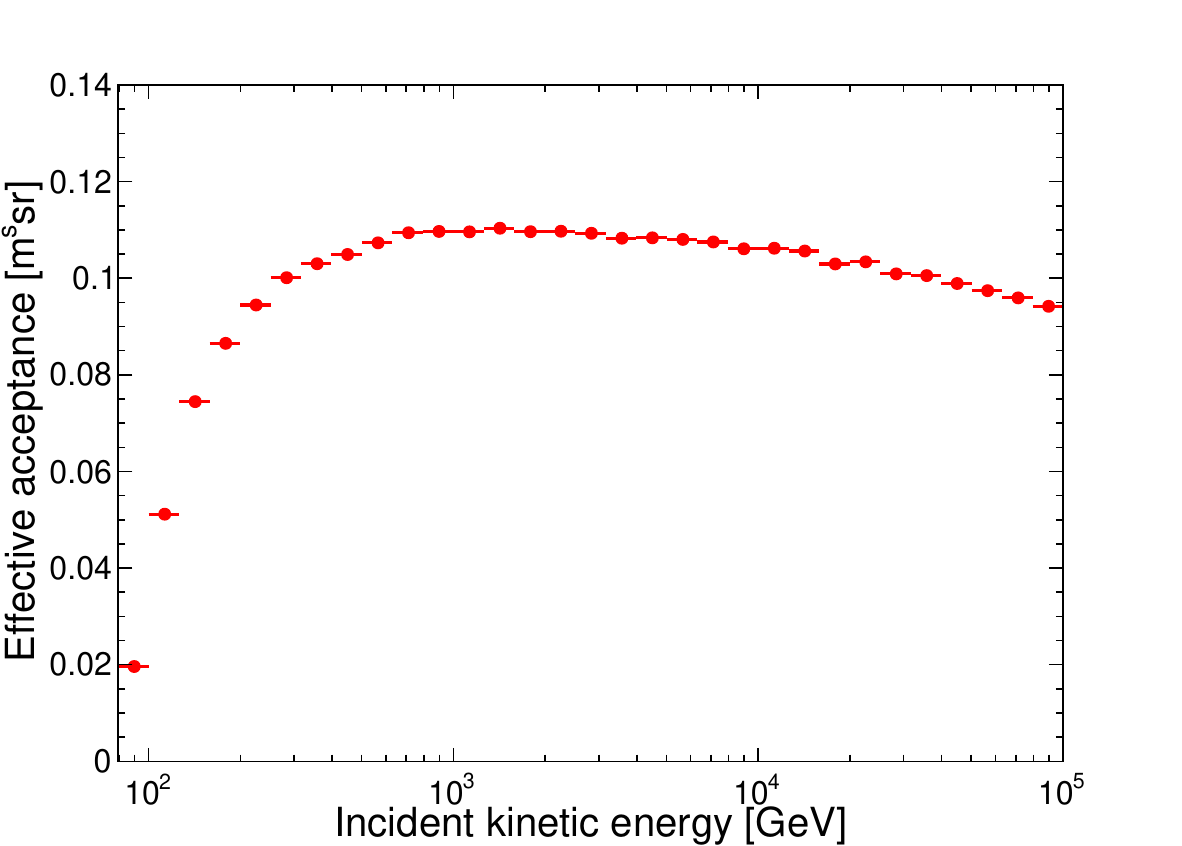}
    \caption{Effective acceptance as computed from B$^{10}$ and B$^{11}$ MC samples for the boron selection.}
    \label{fig:b_acc}
\end{figure}

The flux measurement relies on effective acceptance obtained from the MC simulations, as shown in Fig.~\ref{fig:b_acc}. Therefore it is crucial to validate the simulation by comparing it with the flight data taken from the experiment. To do so, the efficiency of each applied cut is computed with simulation and flight data. To calculate the efficiency with flight data, a carefully selected control sample has to be used. 

\subsubsection{HET efficiency}

\begin{figure}[]
    \centering
    \includegraphics[width=0.45\textwidth]{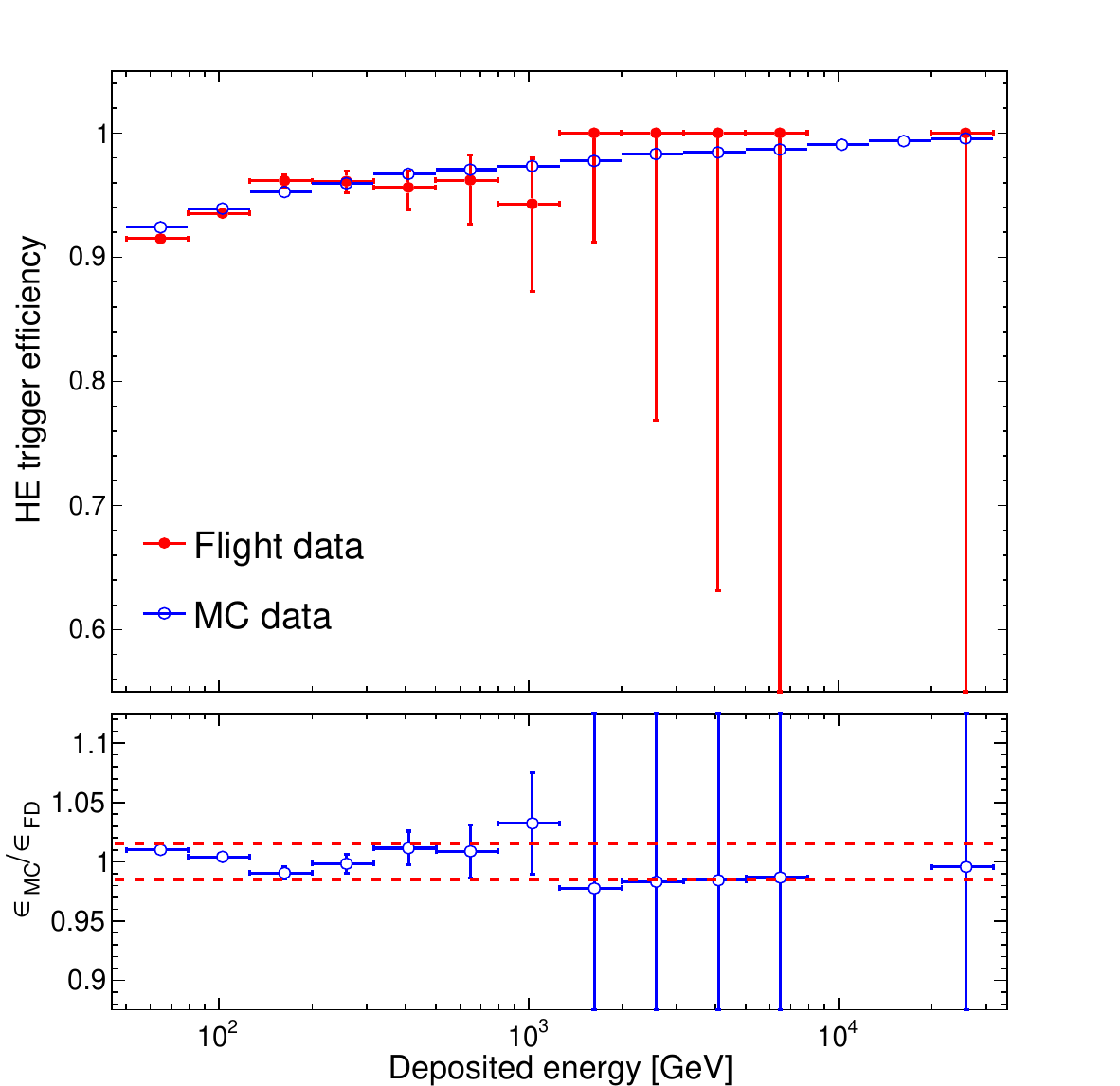}
    \caption{HET efficiency computed for the flight data (blue) and MC simulation (red). The bottom panel shows the ratio, with horizontal dashed lines marking the $\pm 1.5\%$ differences.}
    \label{fig:eff_het}
\end{figure}

To compute the HET efficiency, another trigger defined in the DAMPE DAQ is used. The Low Energy Trigger (LET) has a loose request of an energy deposition of larger than 0.4 MIPs in the first and second layers and larger than 2 MIPs in the third and fourth layers of the BGO. The LET is pre-scaled by a factor of 8 (64) when the satellite is inside (outside) the latitude range of $[-20^{\circ}, 20^{\circ}]$. The efficiency of this trigger for boron is computed to be 100\% with the MC sample. The HET efficiency is computed as
\begin{equation}
    \epsilon_{\rm HET} = \frac{N_{\rm HET|LET}}{N_{\rm LET}},
\end{equation}
where $N_{\rm LET}$ is the number of events passing the full boron sample selection described in this work, but with the request of the LET replacing the HET; $N_{\rm HET|LET}$ is the number of events that pass this selection, plus the HET request.
The efficiencies for the flight data and MC are shown in Fig.~\ref{fig:eff_het}. The uncertainties for the flight data at high energies are relatively large due to the low statistics in the control sample, caused by the LET pre-scaling. The largest discrepancy between the data and MC, taking into account the uncertainties, is 1.5\%, which is assigned as systematic uncertainty due to the HET efficiency.  

\subsubsection{STK track efficiency}

\begin{figure}[]
    \centering
    \includegraphics[width=0.45\textwidth]{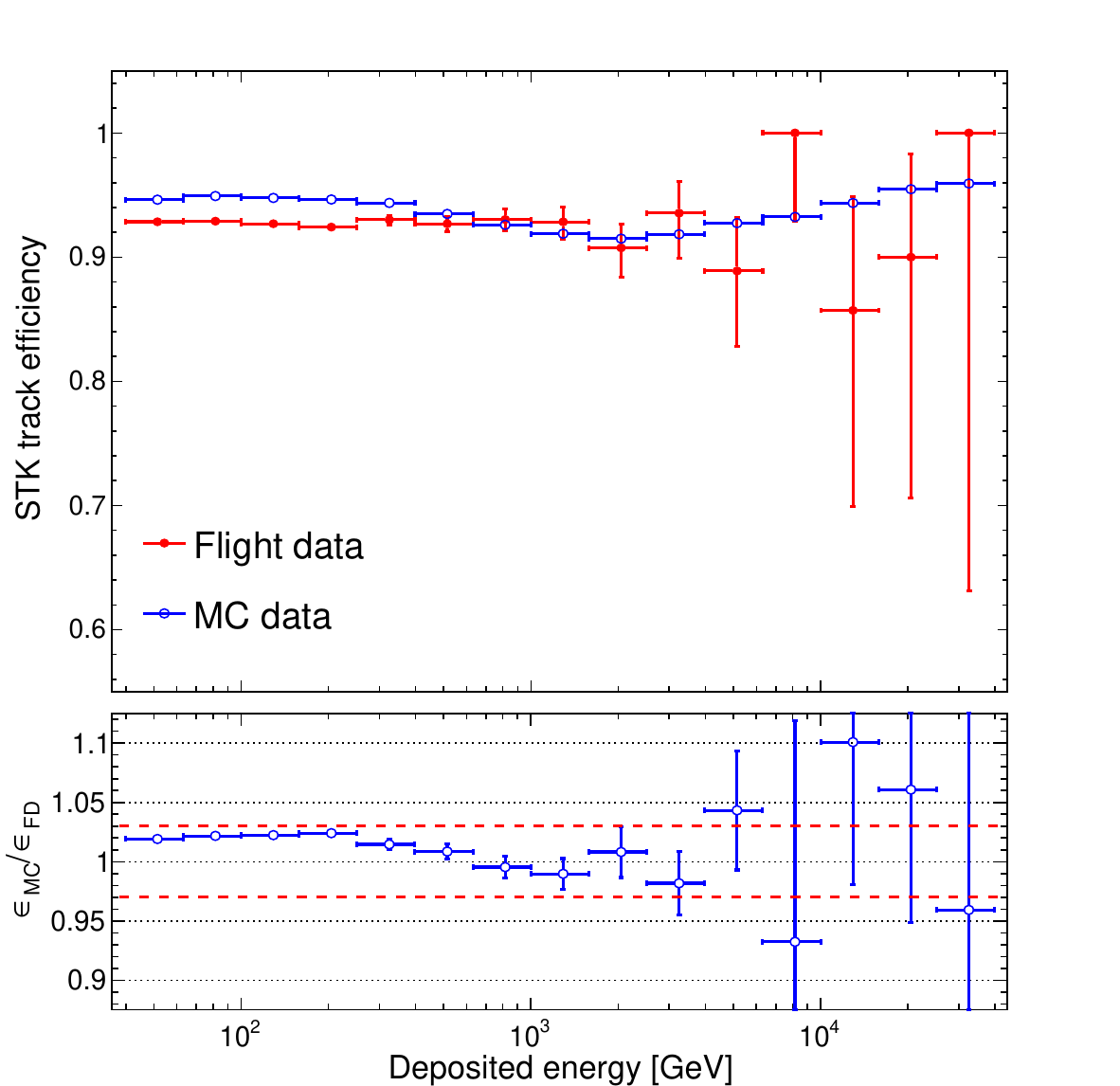}
    \caption{STK track efficiency computed for the flight data (blue) and MC simulation (red). The bottom panel shows the ratio, with horizontal dashed lines marking the $\pm 3\%$ differences.}
    \label{fig:eff_trk}
\end{figure}

To validate the STK track efficiency, a specific boron sample is first selected on the basis of the BGO track, i.e. the shower axis in the BGO calorimeter. Due to the limited angle resolution of the BGO track, a strong correlation with the maximum PSD hit strip is requested in both PSD-x and PSD-y layers. The track efficiency is then estimated as the ratio of the number of events that pass the STK track selection to the total number in such a sample.
\begin{equation}
    \epsilon_{\rm STK-track} = \frac{N_{\rm STK-track|BGO-track}}{N_{\rm BGO-track}},
\end{equation}
The agreement between data and simulation is shown in Fig.~\ref{fig:eff_trk}, with a maximum discrepancy of 3\%, which is assigned as the related systematic uncertainty.

\subsubsection{STK charge selection efficiency}
\begin{figure}[]
    \centering
    \includegraphics[width=0.45\textwidth]{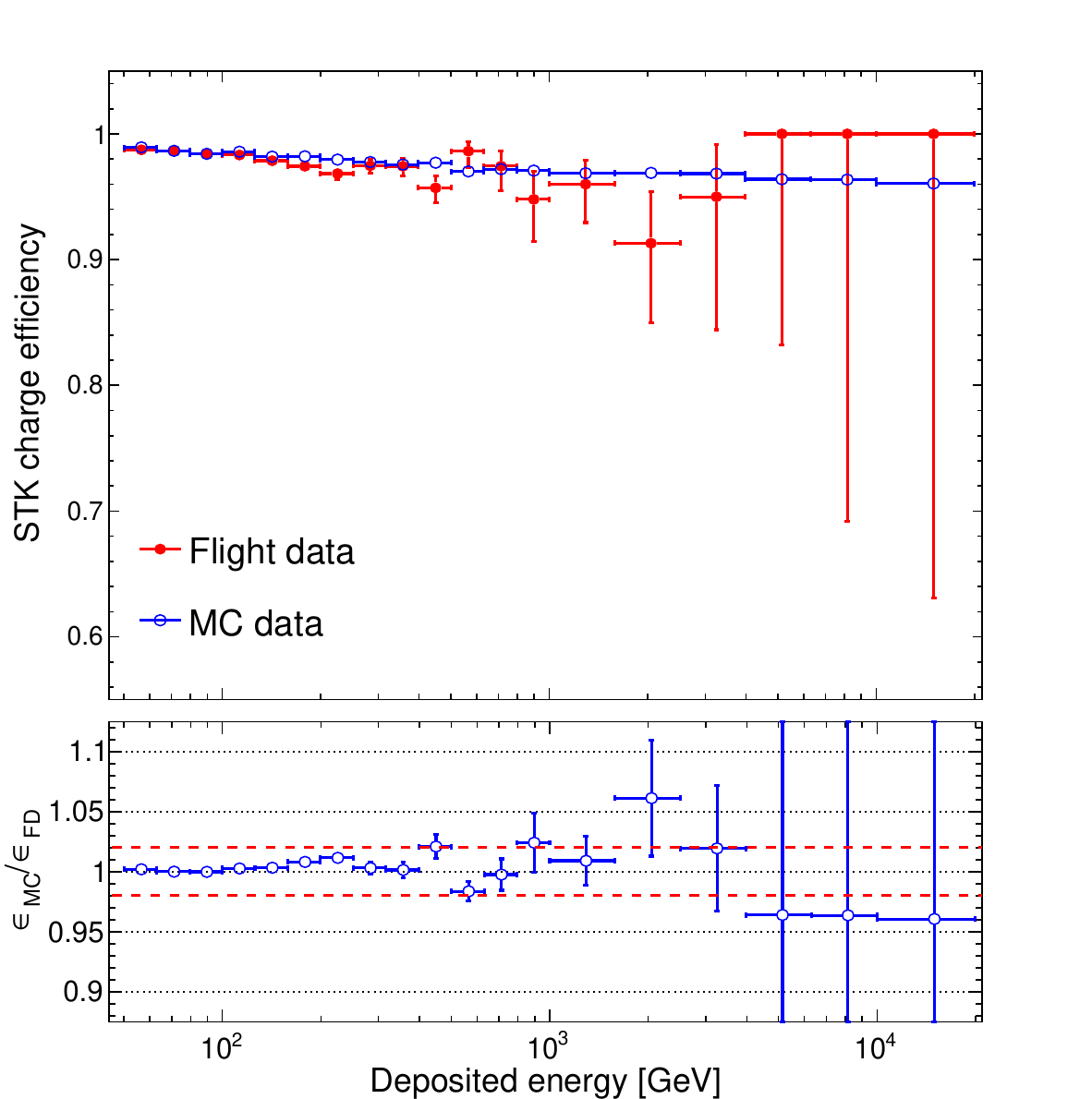}
    \caption{The STK charge selection efficiency computed for the flight data (blue) and MC simulation (red). The bottom panel shows the ratio, with horizontal dashed lines marking the $\pm 2\%$ differences.}
    \label{fig:eff_stkQ}
\end{figure}

The control sample to estimate the STK charge efficiency is chosen using the energy deposited in the second layer of the STK. 
A much more strict PSD charge selection window is chosen to minimize the background from those primary nuclei in the control sample. The STK charge selection efficiency is then computed as
\begin{equation}
    \epsilon_{{\rm STK1}} = \frac{N_{{\rm STK1}|{\rm STK2}|{\rm PSD}}}
    {N_{{\rm STK2}|{\rm PSD}}}.
    \label{eq:STK_eff}
\end{equation}
The agreement between data and simulation is shown in Fig.~\ref{fig:eff_stkQ}, with a maximum discrepancy of 2\%, which is assigned as the related systematic uncertainty.

\subsection{Uncertainty summary}
\begin{figure}[!t]
    \centering
    \includegraphics[width=0.45\textwidth]{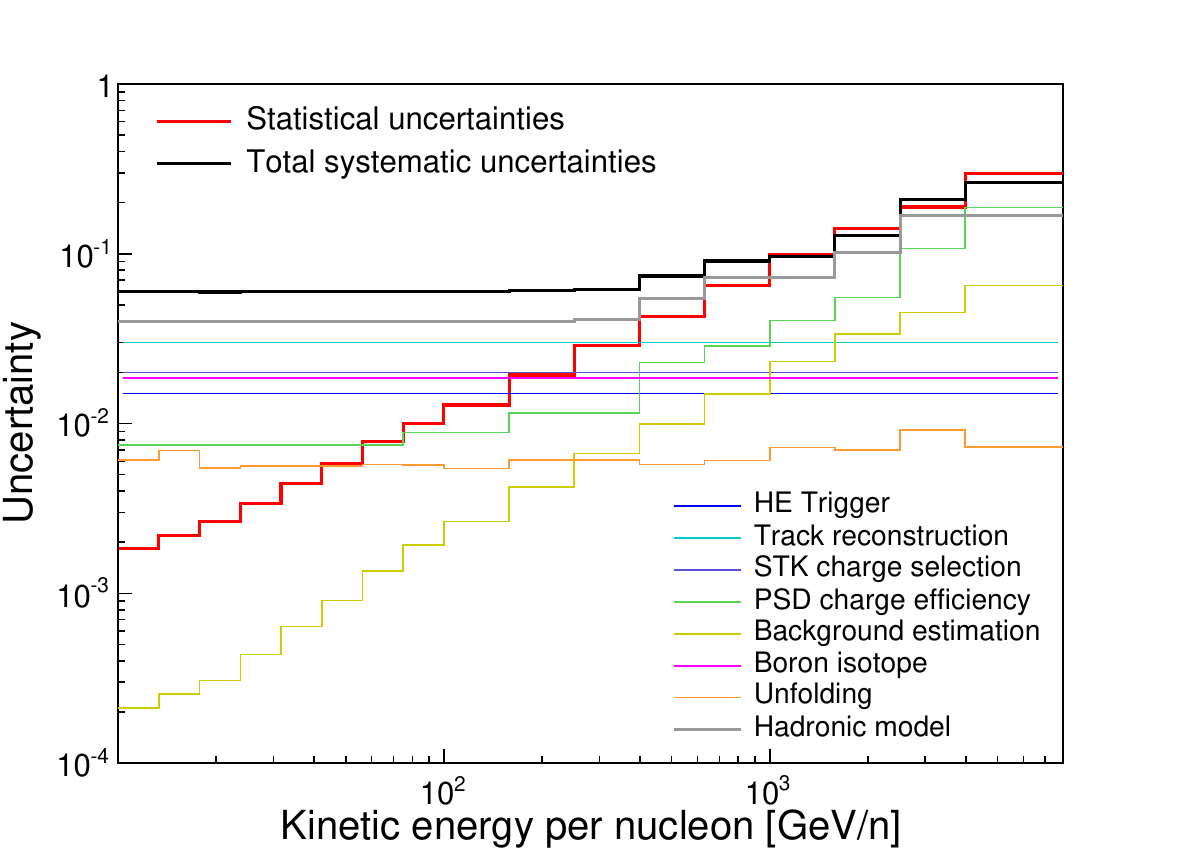}
    \caption{Relative uncertainties on the boron flux measurements from different contributions. The total systematic uncertainties, shown in black, are the quadratic sum of all systematic ontributions.}
    \label{fig:b_statsys_err}
\end{figure}

The total uncertainty budget is shown in Fig.~\ref{fig:b_statsys_err}. Table \ref{Table:b_flux_ek} presents the fluxes and associated uncertainties of boron measured with DAMPE.

\begin{table*}[htbp]
\centering
\caption{The boron fluxes as function of kinetic energy per nucleon measured by DAMPE, together with the 1$\sigma$ statistical errors and the systematic uncertainties from the analysis and hadronic interaction models. The average mass number used to compute the kinetic energy per nucleon is $A=10.7$.}
\begin{tabular}{ c | c | c | c}
\hline
  $E_k$ & $E_{k,{\rm low}}$ & $E_{k,{\rm high}}$ & $\Phi \pm \sigma_{\rm stat}  \pm \sigma_{\rm sys}^{\rm ana} \pm \sigma_{\rm sys}^{\rm had}$ \\
  
  [GeV/n] & [GeV/n] & [GeV/n] & [GeV/n$^{-1}m^{-2}s^{-1}sr^{-1}$]\\
\hline
11.5 & 10.0 & 13.3 & $( 5.25 \pm 0.01 \pm 0.23 \pm 0.21  ) \times 10^{-3} $ \\  
15.3 & 13.3 & 17.8 & $( 2.36 \pm 0.01 \pm 0.11 \pm 0.09) \times 10^{-3}$ \\
20.5 & 17.8 & 23.7 & $( 1.04 \pm 0.01 \pm 0.05 \pm 0.04) \times 10^{-3}$ \\
27.3 & 23.7 & 31.6 & $( 4.43 \pm 0.02 \pm 0.20 \pm 0.18 ) \times 10^{-4} $ \\ 
36.4 & 31.6 & 42.2 & $( 1.86 \pm 0.01 \pm 0.08 \pm 0.07 ) \times 10^{-4}$ \\ 
48.5 & 42.2 & 56.2 & $( 7.77 \pm 0.05 \pm 0.34 \pm 0.31) \times 10^{-5}$ \\
64.7 & 56.2 & 75.0 & $ (3.25 \pm 0.03 \pm 0.14 \pm 0.13 ) \times 10^{-5} $ \\
86.3 & 75.0 & 100.0 & $ (1.37 \pm 0.01 \pm 0.06 \pm 0.06 ) \times 10^{-5} $ \\
124.8 & 100.0 & 158.5 & $( 4.49 \pm 0.06 \pm 0.20 \pm 0.18) \times 10^{-6} $ \\
197.8 & 158.5 & 251.2 & $( 1.18 \pm 0.02 \pm 0.05 \pm 0.05) \times 10^{-6} $ \\
313.5 & 251.2 & 398.1 & $( 3.29 \pm 0.10 \pm 0.15 \pm 0.14) \times 10^{-7} $ \\
496.8 & 398.1 & 631.0 & $( 9.50 \pm 0.41 \pm 0.48 \pm 0.52) \times 10^{-8}$ \\
787.4 & 631.0 & 1000 & $( 2.60 \pm 0.17 \pm 0.14 \pm 0.19) \times 10^{-8}$ \\
1248 & 1000 & 1585 & $(7.18 \pm 0.71 \pm 0.46 \pm 0.52) \times 10^{-9}$  \\
1978 & 1585 & 2512 & $(2.31 \pm 0.32 \pm 0.18 \pm 0.23) \times 10^{-9}$   \\
3135 & 2512 & 3981 & $(7.16 \pm 1.35 \pm 0.89 \pm 1.20) \times 10^{-10}$   \\
5515 & 3981 & 7943 & $(1.67 \pm 0.49 \pm 0.34 \pm 0.28) \times 10^{-10} $ \\ 
\hline
\end{tabular}
\label{Table:b_flux_ek}
\end{table*}

\subsection{Spectral fitting}
The DAMPE boron spectrum is fit with two different models. One is a power-law (PL) function
\begin{equation}
    \Phi_{\rm PL}(E_k) = \Phi_0 \cdot \left( \frac{E_k}{\text{GeV/n}} \right)^{-\gamma},
\label{eq:sbpl}
\end{equation}
and the other is a Smoothly Broken Power Law (SBPL) function
\begin{equation}
    \Phi_{\rm SBPL}(E_k) = \Phi_0 \cdot \left( \frac{E_k}{\text{GeV/n}} \right)^{-\gamma_1} \left[ 1 + \left( \frac{E_k}{E_{\rm br}} \right)^s \right]^{\frac{\Delta \gamma}{s}}.
\label{eq:sbpl}
\end{equation}
The SBPL model has five parameters: the normalization $\Phi_0$, the break energy $E_{\rm br}$, the spectral index before the break $\gamma_1$, the change of spectral index $\Delta \gamma$, and the smoothness parameter of the break $s$. 

To perform the fit, a $\chi^2$ is defined, taking into account the correlation between the points introduced by the unfolding procedure, as
\begin{equation}
\begin{split}
\chi^2 = & \sum_{i}\sum_{j} \left[\Phi(E_k^i)S(E_k^i,w)-\Phi_i \right] \mathcal{C}_{ij}^{-1} \\ & \left[ \Phi(E_k^j)S(E_k^j,w)-\Phi_j \right] + \sum_{k=0}^3 \left( \frac{1-w_k}{\Tilde{\sigma}_{{\rm sys},k}} \right)^2,
\label{eq:chi2_corr}
\end{split}
\end{equation}
where the index $i,\,j$ go over the kinetic energy bins, $\Phi(E_k)$ is the function to model the flux, $\Phi_i$ is the measured flux in the $i$-th bin, $\mathcal{C}_{ij}^{-1}$ is the inverse of the covariance matrix between the $i$-th and $j$-th energy bins, $S(E_k^i,w)$ is a piece-wise function defined as: $S(E_k^i,w)=w_0 ~ \text{if }\log[E_k^i/(\text{GeV/n})] < 1.8$, $w_1~\text{if } 1.8 \le \log[E_k^i/(\text{GeV/n})] < 2.6$, $w_2~\text{if } 2.6 \le \log[E_k^i/(\text{GeV/n})] < 3.2$, and $w_3~\text{if } \log[E_k^i/(\text{GeV/n})] \ge 3.2$. Here $w_j$ are the so-called nuisance parameters defined in different ranges of energy, to take into account the potential bias in the flux measurement introduced by the systematic uncertainties. In this case the energy range of the flux measurement is divided into four sections, each one corresponding to a nuisance parameter. Finally $\Tilde{\sigma}_{{\rm sys},k}$ is the sum of the relative systematic uncertainties. The fit is done for an energy range of 25 GeV/n to 8 TeV/n. Given the relatively large uncertainties at high energies, the smoothness parameter s cannot be effectively constrained, and it is fixed as 5 for a consistency with that adopted in our proton and Helium analyses \cite{p_dampe,he_dampe}. The best fit $\chi^2$ value over the number of degree of freedom is $\chi^2$/d.o.f.$=1.7/10$, for the SBPL model. As a comparison, for the PL model, we get $\chi^2$/d.o.f.$=71.3/12$. This gives a significance of the hardening of 8$\sigma$. The best fit results of the parameters and the uncertainties for the SBPL model are collected in Table \ref{tab:b_fit_params}. 

\begin{table}[htbp]
\centering
\caption{The parameters of the SBPL model obtained from the fit of the boron spectrum. }
\begin{tabular}{ll}
\hline
\hline
Fit energy range  & 25 GeV/n - 8 TeV/n \\ 
Nuisance parameters & 4 \\ 
\hline
$\Phi_0$ (10$^{-9}$ GeV$^{-1}$ m$^{-2}$ s$^{-1}$ sr$^{-1}$) & $9.76\pm 0.37$ \\
$E_{\rm br}$ (GeV/n) & $182 \pm 24 $ \\ 
$\gamma_1$ & 3.02 $\pm$ 0.01 \\ 
$\Delta\gamma$ & 0.31 $\pm$ 0.05 \\ 
$s$ & 5 (fixed) \\
\hline
\end{tabular}
\label{tab:b_fit_params}
\end{table}

\end{document}